\documentclass[aps,pre,reprint,showpacs,superscriptaddress]{revtex4-1}
\usepackage{graphicx}
\usepackage{dcolumn}
\usepackage{bm}
\begin{document}


\title{Identifying the source of super-high energetic electrons in the presence of pre-plasma in laser-matter interaction at relativistic intensities}

\author{D. Wu}
\affiliation{Shanghai Institute of Optics and Fine Mechanics, 
Chinese Academy of Science, Shanghai 201800, China}
\author{S. I. Krasheninnikov}
\affiliation{University of California-San Diego, La Jolla, California, 92093, USA}
\author{S. X. Luan}
\affiliation{Shanghai Institute of Optics and Fine Mechanics, 
Chinese Academy of Science, Shanghai 201800, China}
\author{W. Yu}
\affiliation{Shanghai Institute of Optics and Fine Mechanics, 
Chinese Academy of Science, Shanghai 201800, China}

\date{\today}

\begin{abstract}
The generation of super-high energetic electrons influenced by pre-plasma at relativistic intensity laser-matter interaction is studied in a one-dimensional slab approximation with particle-in-cell simulations. Different pre-plasma scale-lengths of $1\ \mu\text{m}$, $5\ \mu\text{m}$, $10\ \mu\text{m}$ and $15\ \mu\text{m}$ are considered, showing an increase in both particle number and cut-off kinetic energy of electrons with the increase of pre-plasma scale-length, and the cut-off kinetic energy greatly exceeding the corresponding laser ponderomotive energy. 
A two-stage electron acceleration model is proposed to explain the underlying physics. 
The first stage is attributed to the synergetic acceleration by longitudinal electric field and laser pulse, with its efficiency depending on the pre-plasma scale-length. These electrons pre-accelerated in the first stage could build up an intense electrostatic potential barrier with its maximal value several times as large of the initial electron kinetic energy. Part of energetic electrons could be further accelerated by the reflection off the electrostatic potential barrier, with their finial kinetic energies significantly higher than the values pre-accelerated in the first stage.        

\end{abstract}


\maketitle

\section{Introduction}
The influence of pre-plasma in laser-matter interaction at relativistic intensities has attracted great attention from both experimental and theoretical investigations, because of its significant effects on a number of applications, such as laser driven ion acceleration\cite{PhysRevLett.103.024801,PhysRevLett.102.145002,PhysRevLett.103.135001,PhyPla.20.023012,PhyRevE.90.023101,PhysRevLett.115.064801,PhyPla.20.013101}, fast ignition\cite{PhyPla.1.1626,PhyPla.15.056304,PhysRevLett.104.055002,PhysRevLett.108.115004} and bright x/$\gamma$ ray sources\cite{Nat.Phys.7.867,PhyPla.22.080704}, etc. The pre-plasma produced by the intrinsic laser pre-pulse (usually with ns duration) can be as high as $10\ \mu\text{m}$ for the energetic main pulses of energies tens of $\text{kJ}$ with a typical contrast ratio $10^{-5}$. In the fast ignition related experiments with relatively long pulses (with tens of ps duration), high intensity and high power laser, even the contrast ratio can be as high as $10^{-8}$, considerable pre-plasma can still build up in front of a solid density target. The pre-formed plasma always exists in laser-matter interaction at relativistic intensities, thus the laser pre-plasma interaction is inevitable. 

The fast electron generation due to relativistic intensity laser-matter interaction influenced by preformed plasma has been addressed in a number of experimental and theoretical studies\cite{PhysRevLett.104.055002,PhyPla17.060704,Phys.Rev.E.79.066406,PhyPlas.17.023106,PhyPlas.16.103101,Phys.Rev.E.83.046401,PhyPla.20.012703,App.Phys.Lett.102.224101}, suggesting that the presence of pre-plasma can significantly affect the fast electron distributions. Both experiments and numerical simulations have reported an increase of fast electron generation efficiency with the increase of pre-plasma scale-length. 
The recent particle-in-cell simulations\cite{Phys.Rev.E.83.046401} have observed super-high energetic electrons with the cut-off kinetic energy as high as $100\ \text{MeV}$ at laser of intensity $10^{20}\ \text{W}/\text{cm}^2$ and pre-plasma of scale length $10\ \mu\text{m}$. However the underlying physics, i) \textit{the increase in the generation efficiency of energetic electrons with the increase of pre-plasma scale-length}, and ii) \textit{the acceleration mechanism of super-high energetic electrons with kinetic energy greatly exceeding the ponderomotive energy}, is still unclear, which is the aim of this work.

In order to simulate laser-matter interaction with pico-second duration in the presence of large scale pre-plasma, we choose to use one-dimensional (1-D) particle-in-cell (PIC) simulations\cite{PhyPla.20.023012}, because it is computationally cheap. Although multi-dimensional effects, such as laser filamentation and self-focusing\cite{Phys.Fluids.30.526,Phys.Rev.Lett.33.209}, might play roles in these processes, they are neglected in the present work. We focus on the role of pre-plasma in energetic electron beam generation by using systematic particle-in-cell simulations, where the laser is of intensity $10^{20}\ \text{W}/\text{cm}^2$, and different pre-plasma scale-lengths, $1\ \mu\text{m}$, $5\ \mu\text{m}$, $10\ \mu\text{m}$ and $15\ \mu\text{m}$, are considered. 
The questions, i) ``why the generation efficiency of energetic electrons is increasing with the increase of pre-plasma scale-length'', and ii) ``what is underlying acceleration mechanism of super-high energetic electrons with kinetic energy greatly exceeding the ponderomotive energy'', are answered. A two-stage acceleration model is proposed to identify the source of super-high energetic electrons. The first stage is the synergetic acceleration by longitudinal electric field and laser pulses, with its efficiency depending on the pre-plasma scale-length. The second stage is related to the intense electrostatic potential building in front of the target and the accompanying electron reflection by the intense electrostatic potential barrier. 

This paper is arranged as follows: The details of numerical modelling and simulation results are demonstrated in Sec. II. 
The two-stage acceleration model by analysing the simulation results is proposed in Sec. III to explain the impacts of pre-plasma and identify the sources of energetic electrons. In Sec. IV, the acceleration model is further addressed analytically and numerically. The conclusions are given in Sec. V.    

\begin{figure}
\includegraphics[width=8.00cm]{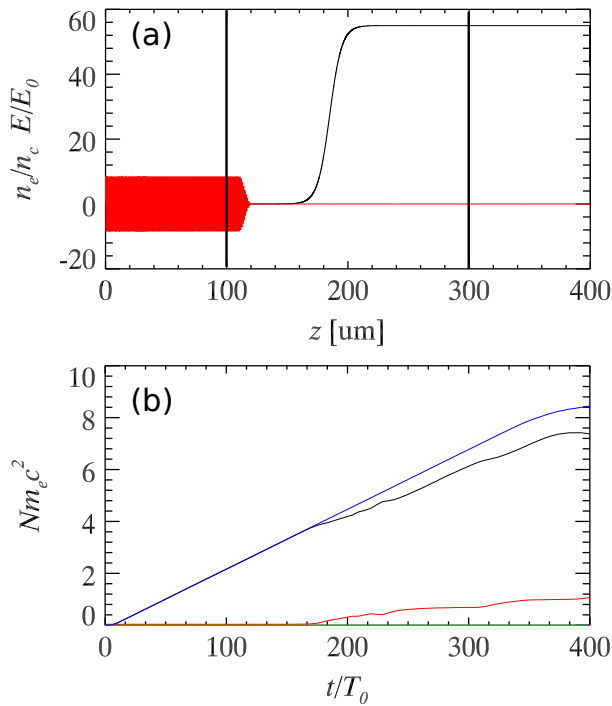}
\caption{\label{fig1} (color online) Schematic of simulation set-up. (a) A linearly polarized laser enters into the simulation box from left boundary and propagates in z-direction. The laser is of intensity $10^{20}\ \text{W}/\text{cm}^2$, where the laser wavelength is $1\ \mu\text{m}$ and pre-plasma scale-length is $10\ \mu\text{m}$. The simulation box is $400\ \mu\text{m}$, and the simulation time is $400T_0$, i.e. $1.3\ \text{ps}$. To analyse the electron energy distributions in detail, we place two diagnostic planes at $z=100\ \mu\text{m}$ and $z=300\ \mu\text{m}$ (shown by the thick black lines), which could time-integrally record the energy distributions of electrons passing through. (b) In order to ensure the accuracy of numerical simulation, we record the temporal variation of laser energy flux (blue line), $\int (E \times B)_z dS dt$, at the left simulation boundary (i.e. $z=0$), the electromagnetic energy (black line), $\int (1/2)(E^2+B^2) dV$, in the simulation box and the particle kinetic energy (red line), $\sum_p m_p(\gamma_p-1)$, in the simulation box. }
\end{figure}

\section{Numerical simulation results} 
The simulations are performed with 1-D PIC code. In order to simulate laser-matter interactions with large scale pre-plasmas, the weighted particle technology is applied in the numerical simulations, which is proven to be more efficient than uniform weighted particles in large density gradients calculations\cite{J.Comp.Phys.227.6846}. In addition, a 4-th order particle cloud and 4-th order FDTD method are applied in our simulations, because these feature makes it suitable for simulating laser solid-density-plasma interactions at relativistic intensities\cite{J.Comp.Phys.227.6846}. The laser is of intensity $10^{20}\ \text{W}/\text{cm}^2$ or normalized amplitude $a=8.54$ (with laser wavelength $1\ \mu\text{m}$), entering the simulation box from the left boundary. The initial plasma density profile is taken as $n_e=n_{\text{solid}}/(1+\exp[-2(z-z_0)/L_p])$, where $n_{\text{solid}}=50n_c$ is the solid plasma density and $L_p$ is the pre-plasma scale-length. As the electron recoiling due to the artificial electrostatic field on the right boundary could interrupt the physics we are studying, to avoid this boundary effect, we choose a large simulation box with the size of $400\ \mu\text{m}$, which is divided into $40000$ cells, with each cell containing $1000$ electrons and $1000$ ions. In our simulations, the region $0<z<100\ \mu\text{m}$ is left as vacuum, $L_p$ varies from $L_p=1\ \mu\text{m}$, $5\ \mu\text{m}$, $10\ \mu\text{m}$ to $15\ \mu\text{m}$, $z_0$ is fixed as $180\ \mu\text{m}$ and the minimum plasmas density is set as $0.001n_c$ for all simulation cases. In order to analyse the electron energy distributions in detail, we have placed two diagnostic planes to temporally record the electrons passing through. As shown in Fig.\ \ref{fig1} (a), the first diagnostic plane is located at $z=100\ \mu\text{m}$ to record the electron going through in -z-direction, and the other one is located at $z=300\ \mu\text{m}$, recording the electron passing through in z-direction.

\begin{figure}
\includegraphics[width=8.00cm]{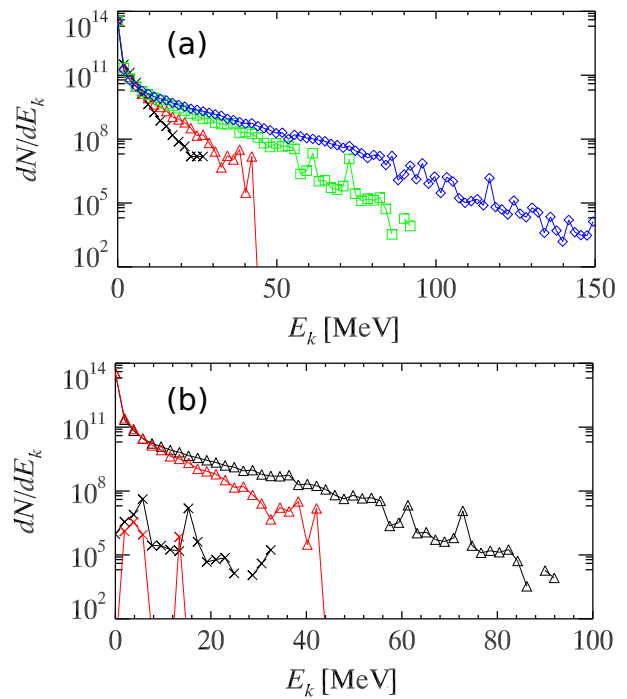}
\caption{\label{fig2} (color online) The laser is of intensity $10^{20}\ \text{W}/\text{cm}^2$, and laser wavelength is $1\ \mu\text{m}$. (a) Electron energy spectra recorded at $z=300\ \mu\text{m}$ at the finial time of simulations. Black line records the energy spectra for pre-plasma of scale-length $1\ \mu\text{m}$, red line is the case for pre-plasma of scale-length $5\ \mu\text{m}$, green line is of scale-length $10\ \mu\text{m}$ and blue line is of $15\ \mu\text{m}$. (b) The black (red) crosses show the spectra of electrons (passing through in -z-direction) recorded by the diagnostic plane located at $z=100\ \mu\text{m}$ and the black (red) triangles show the energy spectra of electrons (in z-direction) recorded by the diagnostic plane located at $z=300\ \mu\text{m}$ with pre-plasma of scale-length $10\ \mu\text{m}$ ($5\ \mu\text{m}$).}
\end{figure}

To ensure the accuracy of the simulation, as we have done previously\cite{PhyPlas.22.093108}, we record the energy history of laser flux energy entering the simulation box ($E_{l}$, blue line), 
electromagnetic field energy in the simulation box ($E_{em}$, black line), and particle kinetic energy in the simulation box ($E_{k}$, red line), which is shown in Fig.\ \ref{fig1} (b). It is clearly demonstrated that, at $t=180T_0$, part of laser flux energy is starting to convert to plasmas. The energy conservation condition $E_{l}=E_{em}+E_{k}$ is always satisfied in the whole simulation. The total simulation time is set to be $400T_0$, to avoid the electron recoiling effect. 

The fast electron energy spectra obtained for different pre-plasma scale-length ($L_p=1\ \mu\text{m}$, $5\ \mu\text{m}$, $10\ \mu\text{m}$ and $15\ \mu\text{m}$) while keeping the laser of intensity $10^{20}\ \text{W}/\text{cm}^2$ fixed, are analysed in Fig.\ \ref{fig2}. The dependence of electron energy distributions on the pre-plasma scale-length is plotted in Fig.\ \ref{fig2} (a), which records the energy spectra of electrons passing through the diagnostic plane located at $z=300\ \mu\text{m}$. There is a clear relation between cut-off kinetic energy and pre-plasma scale-length, the larger the scale-length the higher the cut-off kinetic energy, which is in agreement with earlier published works\cite{Phys.Rev.E.83.046401}. We have also found that the cut-off electron kinetic energy greatly exceeds the corresponding laser ponderomotive energy, which is $3.8\ \text{MeV}$ at intensity $10^{20}\ \text{W}/\text{cm}^2$. For pre-plasma of scale-length $1\ \mu\text{m}$, $5\ \mu\text{m}$, $10\ \mu\text{m}$ and $15\ \mu\text{m}$, the corresponding cut-off energies are $28\ \text{MeV}$, $44\ \text{MeV}$, $92\ \text{MeV}$ and exceeding $150\ \text{MeV}$, respectively. In Fig.\ \ref{fig2} (b), we pick up two cases with pre-plasma of scale-length $5\ \mu\text{m}$ and $10\ \mu\text{m}$, and include the energy spectra of electrons recorded by the diagnostic plane located at $z=100\ \mu\text{m}$.
By comparing the two energy spectra recorded by two different diagnostic planes, we can find that the cut-off energy recorded at $z=300\ \mu\text{m}$ is about three times of the value recorded at $z=100\ \mu\text{m}$. For pre-plasma of scale-length $5\ \mu\text{m}$, the cut-off energy recorded at $z=100\ \mu\text{m}$ is $15\ \text{MeV}$, while that recorded at $z=300\ \mu\text{m}$ is $44\ \text{MeV}$. In addition, for scale-length $10\ \mu\text{m}$, the cut-off energy recorded at $z=100\ \mu\text{m}$ is $32\ \text{MeV}$, while that recorded at $z=300\ \mu\text{m}$ is $92\ \text{MeV}$.
 
\begin{figure*}
\includegraphics[width=16.5cm]{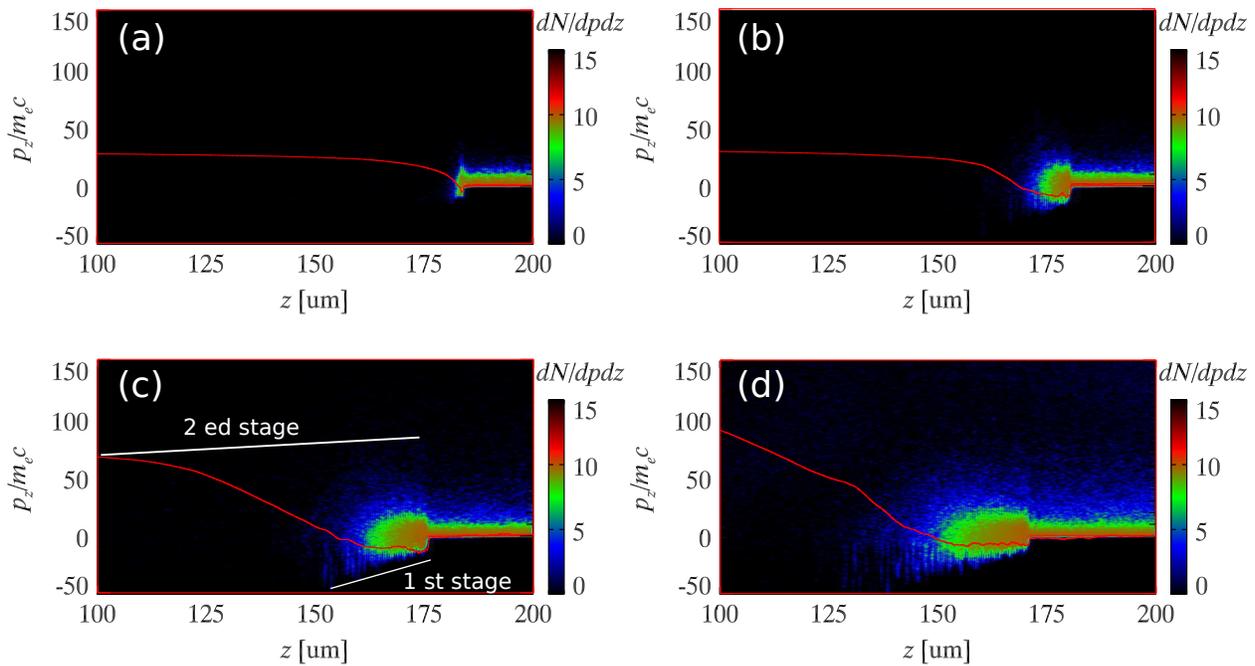}
\caption{\label{fig3} (color online) The laser is of intensity $10^{20}\ \text{W}/\text{cm}^2$, and laser wavelength is $1\ \mu\text{m}$. Comparisons of $z$-$p_z$ phase-space plots with different pre-plasma scale-length, (a) $1\ \mu\text{m}$, (b) $5\ \mu\text{m}$, (c) $10\ \mu\text{m}$ and (d) $15\ \mu\text{m}$, respectively. The red lines covered on the phase-space plots are the electrostatic potential curves ($\int^{z}E_z dz$), normalized by $-e\phi/m_ec^2$. 
Note that the phase-space mixing region and the maximal value of the electrostatic potential barrier increase with the increase of the pre-plasma scale-length. 
The first stage is due to the synergetic acceleration by the longitudinal electric field $E_z$ and the ponderomotive force of the reflected laser pulse, and the second stage is attributed to the intense electrostatic potential building and the accompanying reflection of the energetic electrons off the potential barrier.}
\end{figure*} 

\section{Explaining of simulation results} 
We have found that the cut-off kinetic energy of electrons increases with the increase of the pre-plasma scale-length. 
In the meanwhile, we have also noticed that the cut-off electron kinetic energy recorded by diagnostic plane located at $z=300\ \mu\text{m}$ is three times more or less that recorded at $z=100\ \mu\text{m}$. 
The aim of this work is to uncover the mysteries, i) the increase in the generation efficiency of energetic electrons with the increase of pre-plasma scale-length and ii) the source of super-high energetic electrons with energy greatly exceeding the corresponding laser ponderomotive energy. In order to understand the underlying physics of the observed phoneme, we now refer to analysing the $z$-$p_z$ phase-space dynamics. 
Fig.\ \ref{fig3} describes the phase-space patterns of laser pre-plasma interactions with laser of intensity $10^{20}\ \text{W}/\text{cm}^2$ and pre-plasma of scale-length $L_p=1\ \mu\text{m}$ [Fig.\ \ref{fig3} (a)], $5\ \mu\text{m}$ [Fig.\ \ref{fig3} (b)], $10\ \mu\text{m}$ [Fig.\ \ref{fig3} (c)] and $15\ \mu\text{m}$ [Fig.\ \ref{fig3} (d)], respectively. The phase-space density $D(z, p_z)$ gives a value proportional to the number of electrons found between $z$ and $z+dz$ having longitudinal momentum ranged between $p_z$ and $p_z+dp_z$. The normalized electrostatic potential, $-e\phi/m_ec^2$, due to the longitudinal charge separation field $E_z$, is shown in red curves covered on phase plots. The electron longitudinal momentum $p_z$ is in the dimensionless units of $\gamma \beta$ and $z$ is in the units of laser wavelength, which is $1\ \mu\text{m}$. 

In the very earlier stage of the laser propagation in under-dense preformed plasma, 
part of electrons are swept away in the forward direction by the laser ponderomotive force, leaving behind immobile ions. 
The electric field $E_z$ due to charge separation within the under-dense plasma region tries to pull the electrons in the backward direction. When the laser arrives at the critical density surface and is reflected back, the ponderomotive force of the reflected laser pulse can further accelerate the electrons in the backward direction. Actually, the first stage acceleration is due to the synergetic effects by this longitudinal charge separation field $E_z$ and the ponderomotive force of the reflected laser pulse. 
From Woodward-Lawson theorem\cite{IEE.T.N.S.26.4217}, we know that a single electron in vacuum, oscillating coherently with a propagating plane laser pulse would gain zero cycle averaged energy since the electron energy gain in one half cycle is exactly equal to the energy loss in the next half cycle. 
However, when there exists an external electric field\cite{Phys.Rev.E.83.046401,PhyPla.19.060703,Phy.Rev.Lett.111.065002,PhyPla.21.104510}, even though this field is very week, the Woodward-Lawson theorem can be broken and the electron can obtain none zero energy from the synergetic effects by the external electric field and the laser pulse. 

When the incident laser arrives at the critical density surface and is reflected back, a strong delta-like charge separation field or the step-like electrostatic potential, as shown in Fig.\ \ref{fig3}, is build up therein, which is strong enough to drive electrons to very high velocity within very short time and short length. Imagine we are standing on the frame of a backward propagating electron, we will find that the incident laser pulse is oscillating very fast, and its only contribution to the motions of the electron is to increase its mass by a factor $\gamma=(1+a^2/2)^{1/2}$ in an average way (Appendix A), however the reflected laser pulse is oscillating so slow that this electron can be captured and continually be accelerated backward by its ponderomotive force. 
Actually the first stage acceleration strongly depends on the pre-plasma scale-length. As clearly demonstrated in Fig.\ \ref{fig3} (c) [$10\ \mu\text{m}$], the first stage acceleration is stronger than that in Fig.\ \ref{fig3} (a)[$1\ \mu\text{m}$] and (b) [$5\ \mu\text{m}$], but not as efficient as that in Fig.\ \ref{fig3} (d) [$15\ \mu\text{m}$]. 
According to Woodward-Lawson theorem\cite{IEE.T.N.S.26.4217}, a single electron can not gain none zero cycle-averaged energy from one plane wave. However, in our case, there exists an external electric field $E_z$ due to the charge separation field in the under-dense pre-plasma region. Actually, as we shall analyse in the next section, the pre-plasma scale-length determines the space extension of $E_z$, which eventually determines the maximum possible electron energy gain in this synergetic acceleration process.

The energetic electrons pre-accelerated in the first stage continuously propagate backward and expand freely, building up an intense electrostatic potential barrier therein, as shown by the red curves in Fig.\ \ref{fig3}. 
Actually the peak value of the electrostatic potential barrier is three times as large of the kinetic energy of these electrons pre-accelerated. However at present, the claiming ``three times'' only have statistical meanings. 
As we know, for an electron with kinetic energy $E_{k\text{in}}$ initially located at position with zero electrostatic potential energy, it is impossible to arrive at the position with potential energy $U_p>E_{k\text{in}}$ without any external forces. However, for the continuously emitting electron beams or separated multi electron bunches, we find that part of electrons can arrive at positions where the potential energies are several times as large of their initial kinetic energies. 
When these electrons are reflected back to their original positions, the obtained kinetic energies of the returned electrons will increase to $E_{k\text{f}\max}=N\times E_{k\text{in}}$. Although it seems impossible, this process conserves the total energies of the system, and $\sum n_{\text{in}} E_{k\text{in}}=\sum n_{\text{f}} E_{k\text{f}}$ is always satisfied, with $E_{k\text{f}}$ having $E_{k \text{f}\min}<E_{k\text{in}}<E_{k\text{f}\max}$. In the next section, solid interpretations are presented, including analytical analysis and electrostatic numerical simulations, for the building process of electrostatic potential and the accompanying electron kinetic enhancement by the reflection off this potential barrier.  

\section{Two-stage acceleration model} 

\textbf{\textit{The synergetic acceleration by longitudinal electric field and laser ponderomotive force--}} We consider the relativistic electron dynamics in the presence of two counter-propagating plane laser waves with vector potential $a_{+}$ and $a_{-}$ and longitudinal field $E_z$. $a_{+}$ means the propagating of laser pulse is with the same propagation direction of electron in the presence of electric field $E_z$. Considering the electron propagates with high velocity along the z-direction, the only contribution of the incident wave $a_{-}$ is to increase the electron mass in an averaged way. The z-momentum and energy equation, in normalized units, can be written as  

\begin{equation}
\label{eom_pz}
\frac{d(\gamma v_z)}{dt}=\frac{-1}{2\gamma}\frac{\partial a_{+}^2}{\partial z}+E_z,
\end{equation}
\begin{equation}
\label{eom_energy}
\frac{d \gamma}{d t}=\frac{1}{2\gamma}\frac{\partial a_{+}^2}{\partial t}+E_z v_z,
\end{equation}
where $v_z$ is the electron velocity component along z-direction and the relativistic factor $\gamma$ defined as $\gamma=\gamma_{A}\gamma_{z}$ with $\gamma_{A}=(1+a^2/2+a_{+}^2)^{1/2}$, $a^2/2$ is the average mass increase due to the incident laser wave of the form $a_{-}=a\sin(t+x)$, and $\gamma_{z}=1/(1-v_z^2)^{1/2}$. 

\begin{figure}
\includegraphics[width=8.0cm]{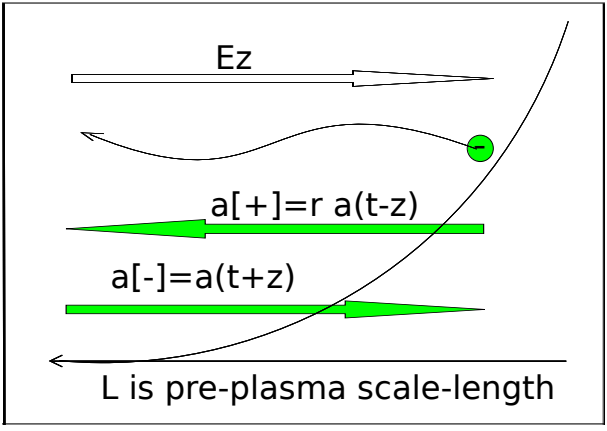}
\caption{\label{fig4} (color online) Schematic demonstration of the electron dynamics in the presence of two counter-propagating laser waves and longitudinal electric field $E_z$. Note that the $a_{+}$ means the propagating of laser pulse is with the same propagation direction of electron in the presence of electric field $E_z$, where $r^2$ is the reflection rate compared with the incident laser $a_{-}$.}
\end{figure}

\begin{figure}
\includegraphics[width=8.0cm]{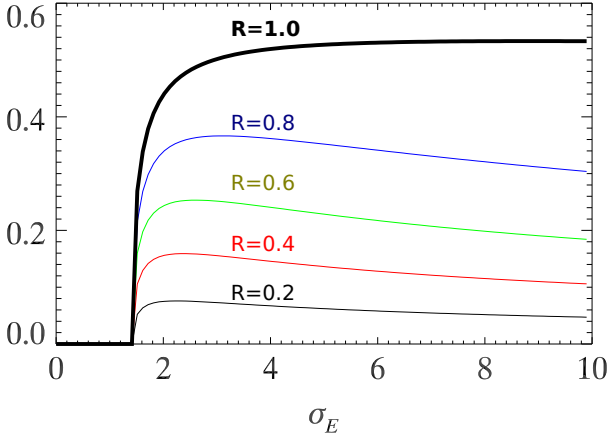}
\caption{\label{fig5} (color online) Factor $\eta$ as function of $\sigma_E$ and $R$.}
\end{figure}

For reflecting plane wave of the form $a_{+}=a_{+}\sin(t-z)$, from Eqs. (\ref{eom_pz}) and (\ref{eom_energy}), we find 
\begin{equation}
\label{eom_pz_energy}
\frac{d}{dt}\gamma_A\gamma_z(1-v_z)=-E_z(1-v_z).
\end{equation}

For constant electric field, Eq. (\ref{eom_pz_energy}) can be integrated and we have 
\begin{equation}
\label{integral}
\gamma_A\gamma_z(1-vz)=\sigma_{\tau_0}-E_z(t-t_0-z-z_0),
\end{equation} 
where $t_0$ is the time at which the electron crosses $z=z_0$ and $\sigma_{\tau_0}=\gamma_A\gamma_z(1-v_z)|_{t=t_0,z=z_0}$. 
Note for the highly relativistic case, we have $\sigma_{\tau_0}\sim (1/2)(\gamma_A/\gamma_z) \ll 1$.

The trajectory of the electron $z$ can be found by introducing a local time $\tau=t-z$, 
in which $d\tau/d\tau=dt/d\tau-dz/d\tau$ and $dt/d\tau=(dz/dt)(dt/d\tau)/v_z$, as 
\begin{equation}
\label{dz_simple}
\frac{dz}{d\tau}=\frac{v_z}{1-v_z}.
\end{equation}
Using $v_z$ from Eq. (\ref{integral}), $dz/d\tau$ can be found to be
\begin{equation}
\label{dz}
\frac{dz}{d\tau}=\frac{1}{2}[f^2(\tau)-1],
\end{equation}
where $f(\tau)=\gamma_A(\tau+\tau_0)/(\sigma_{\tau_0}-E_z\tau)$.

The change in the electron energy only due to the contribution of laser waves, $\sigma \varepsilon(\tau)$ is given by 
$\Delta \varepsilon(\tau)=\gamma_A(\tau+\tau_0)\gamma_z(\tau+\tau_0)-\gamma_A(\tau_0)\gamma_z(\tau_0)-E_z[z(\tau+\tau_0)-z(\tau_0)]$. Following the Eq. (\ref{integral}) and making use of the inequality ($\sigma_{\tau_0} \ll 1$, $\sigma_{\tau+\tau_0} \ll 1$ and $E_z\tau \ll 1$), $\Delta \varepsilon(\tau)$ can then be rewritten as 
\begin{equation}
\label{dE}
\Delta \varepsilon(\tau)=\frac{1}{2}\int_{0}^{\tau}{\frac{d\gamma_A^2(\tau+\tau_0)/d\tau}{\sigma_{\tau_0}-E_z\tau}}d\tau
\end{equation} 

Through Eqs. (\ref{dz}) and (\ref{dE}), we can find the maximal-possible energy gain within the limited longitudinal scale length $L$ and the maximal in-phase time $\tau=\pi/2$,
\begin{equation}
\label{L}
L=\frac{1}{2E_z^2}[\frac{\gamma_A^2(\pi/2+\tau_0)}{\sigma_{E}-\pi/2}-\frac{\gamma_A^2(\tau_0)}{\sigma_E}
  -a_{+}^2f(\sigma_E)]-\frac{\pi}{4}, \nonumber
\end{equation}
\begin{equation}
\label{Energy}
\Delta \varepsilon(\pi/2)=\frac{a_{+}^2}{2E_z}f(\sigma_E),
\end{equation}
where we define $\sigma_E=\sigma_{\tau_0}/E_z \geq \pi/2$, and
\begin{equation}
\label{f}
f(\sigma_E)=\int_{0}^{\pi/2}\frac{\sin{(2x)}}{\sigma_E-x}dx.
\end{equation}

As $\tau_0$ is just an arbitrary initial local time, for simplicity we set $\tau_0=0$ in the following expressions. Assuming $a \gg 1$, $L \gg 1$ and $a_{+}^2=Ra^2$, where $R$ is the reflection rate, based on Eq. (\ref{L}) we can obtain, 
\begin{equation}
\label{E}
E_z=\frac{a}{L^{1/2}}\sqrt{[\frac{R\sigma_E+\pi/4}{2\sigma_E(\sigma_E-\pi/2)}-\frac{R}{2}f(\sigma_E)]}.
\end{equation}

Combining Eq. (\ref{Energy}) and Eq. (\ref{E}), the maximal-possible electron kinetic energy gain within the limited longitudinal length $L$ from the laser of incident amplitude $a$ and reflection rate $R$ can be expressed as,
\begin{equation}
\label{Scaling}
\Delta \varepsilon=\eta a L^{1/2}=\frac{R}{2}\frac{f(\sigma_E)}{g(\sigma_E)} a L^{1/2},
\end{equation}
with $g^2(\sigma_E)=(R\sigma_E+\pi/4)/[2\sigma_E(\sigma_E-\pi/2)]-{Rf(\sigma_E)}/{2}$.

In Eq. (\ref{Scaling}), the coefficient $\eta$ is the function of $R$ and $\sigma_E$. From Fig.\ \ref{fig5}, for the typical reflection rate $R=0.9$, $\alpha$ almost saturates at $0.5$ for a large range of $\sigma_E$. Finally, we give a scaling law which describes the maximal-possible electron energy gain for the synergetic acceleration process, where the laser intensity $I$ is normalized by $1.37\times10^{18}\ \text{W}/\text{cm}^2$ and the longitudinal length $L\sim\beta L_p$ is normalized by $\mu\text{m}$,
\begin{equation}
\label{scaling-law}
\varepsilon\ [\text{MeV}]=0.64\times \beta^{1/2} \times I^{1/2} \times L_p^{1/2}.
\end{equation} 

In Eq. (\ref{scaling-law}), we assume that the longitudinal length is on the order of pre-plasma scale-length with $L\sim\beta L_p$. Here we give an estimated value of $\beta$ as $2.5$, by comparing the actual longitudinal acceleration extension $L\sim25\ \mu\text{m}$ and pre-plasma scale-length $L_p=10\ \mu\text{m}$ in Fig.\ \ref{fig3} (c) and the actual longitudinal acceleration extension $L\sim40\ \mu\text{m}$ and pre-plasma scale-length $L_p=15\ \mu\text{m}$ in Fig.\ \ref{fig3} (d). 
According to the scaling law of Eq.\ (\ref{scaling-law}), we can see that the first stage acceleration, or the synergetic acceleration by longitudinal electric field $E_z$ and the ponderomotive force of the reflected laser, depends on both the incident laser intensity and the pre-plasma scale-length.

\textbf{\textit{Electrostatic potential building and the accompanying electron reflection--}} To get the insights on both i) the possibility of the formation of the electrostatic potential barrier with the maximal value significantly larger than electron kinetic energy, and ii) the role of the potential barrier in electron acceleration, let us consider 1-D model problem. Assume that at $t=0$ we have a bunch of electrons with density $n_b$ occupied region $0<z<z_b$ ($z_b\ll\lambda_{De}$) with momentum $p_0>0$ and a bunch of immobile ions, located at $z<0$ such that total electron and ion charges compensate each other. We consider dynamics of electron bunch expansion assuming that the electrons, which come back to their original positions, do not move any-more. Since we are considering the 1-D geometry, then the electric field acting on electron is solely depends on its original position at $t=0$ and does not vary in time. Therefore, for the electron having $z(t=0)=z_0<z_b$ we have the following equation of motion,
\begin{equation}
\label{eom_bunch}
\frac{d}{dt}\frac{p}{\sqrt{1-p^2}}=-E_z(z_0),
\end{equation}
where $E_z(z_0)$ is the original electric field which is normalized by $e/m_ec$. From Eq.\ (\ref{eom_bunch}) we find the time dependence of the position $z(t,z_0)$ of the electron initially located at $z_0$ as
\begin{widetext}
\begin{equation}
\label{eom_z}
z(t,z_0)=z_0+\int_0^t\frac{p_0-E_z(z_0)t^{'}}{\sqrt{1+[p_0-E_z(z_0)t^{'}]^2}}dt^{'}
=z_0-\frac{1}{E_z(z_0)}\{\sqrt{1+[p_0-E_z(z_0)t]^2}-{\sqrt{1+p_0^2}}\},
\end{equation}
\end{widetext}
where $p_0=p(t=0)$. From Eqs.\ (\ref{eom_bunch}) and (\ref{eom_z}) one can easily see that within the setting of the problem the
electrons coming back to its original position have $p=-p_0$ and, therefore, acquire the original energy.

The original increase of the normalized electrostatic potential within the electron bunch, $\delta\phi_0$, can be easily found from Poisson equation,
\begin{equation}
\label{poisson}
\delta\phi_0=\frac{1}{2}(\frac{\omega_{pe}z_b}{c})^2,
\end{equation}
where $\omega_{pe}^2= 4\pi e^2n_b/m$. Now we will analyse time variation of the electrostatic potential at
relatively large time $t>p_0/E_z(z_0)$, when the majority of electrons already came back to their original
positions. Estimating the magnitude of $E_z(z_0)$ from the Poisson equation, we can re-write this inequality as,
\begin{equation}
\label{tau}
t>\tau_{b}=\frac{p_0c}{\omega_{pe}^2z_b}.
\end{equation}
Then the difference of the normalized electrostatic potential, $\Delta\phi(t)$, 
between the head of expanding electron bunch, $z_h(t)=z(t,z_b)$, and the coordinate $z_r(t)$ with $z_r=z(t, z_r)$ of electrons
returning to its original position at time $t$, can be written as follows,
\begin{equation}
\Delta\phi(t)=\int_{z_r(t)}^{z_h(t)}E_z(z)dz, \nonumber
\end{equation}
or, 
\begin{widetext}
\begin{equation}
\label{deltafi}
\Delta\phi(t)=-\int_0^{E_z[z_r(t)]}E_z(z_0)\frac{dz(t,z_0)}{dE_z(z_0)}dE_z(z_0) \\
=-\frac{1}{2}(\frac{\omega_{pe}}{c})^2[z_b^2-z_r^2(t)]+\int_0^{2p_0}\frac{\sqrt{1+p_0^2}-\sqrt{1+(p_0-\xi)^2}}{\xi}d\xi.
\end{equation}
\end{widetext}
Since we are considering the time $t\gg\tau_b$ where $z_r(t)\rightarrow z_b$, we find the following asymptotic expression, 
$\Delta\phi_{\infty}=\Delta\phi(t\rightarrow \infty)$,
\begin{equation}
\label{fiinfinity}
\Delta\phi_{\infty}= \int_0^{2p_0}[\sqrt{1+p_0^2}-\sqrt{1+(p_0-\xi)^2}]\frac{d\xi}{\xi}.
\end{equation}
From Eq.\ (\ref{fiinfinity}) we derive $\Delta\phi_{\infty}\sim p_0^2$ for $p_0\ll1$ and $\Delta\phi_{\infty}\sim 2\ln(2)p_0^2$ for $p_0\gg1$. In other words, for non-relativistic case $\Delta\phi_{\infty}$ is twice of the initial electron kinetic energy $E_{k\text{in}}$, while for a super-relativistic case $\Delta\phi_{\infty}\sim2\ln(2)E_{k\text{in}}\sim1.4E_{k\text{in}}$.

As we mentioned before, electrons, being finally reflected back by potential, will come to their original positions and obtain their original kinetic energy. So that in the process of launching just one electron bunch, 
there is no possibility of increasing electron energy. 
However, situation changes drastically, when we launch a few electron bunches separated by a dwell time $\tau_{\text{dw}}$. 
To get an insight in electron acceleration mechanism, consider the case of two bunches. 
The first bunch, launched at $t=0$ will expand as it was discussed before. At time $t=\tau_{\text{dw}}>\tau_b$, the second bunch starts launching. At that time, the first bunch has formed the ``potential barrier'' between the head of the first bunch and the launch point, with $\phi_{\text{bar}}=\Delta\phi_{\infty}$. However, almost all electrons of the first bunch already came back to their original position and the electric field within the ``potential barrier'' becomes very small,
with $E\sim\Delta\phi_{\infty}/p_0t\ll E_z(z_0)$. 
As a result, second bunch also expands virtually into vacuum and at the time $t=2\tau_{\text{dw}}$, 
the cumulative contribution of the first and second bunches will create the ``potential barrier'' with $\phi_{\text{bar}}=2\times\Delta\phi_{\infty}$. In addition, relatively small amount of electrons at the head of the first bunch turning back after the expansion of the second bunch will finally acquire not only their initial kinetic energy but also potential energy created by the second bunch. As a result, their total kinetic energy when they are reaching the launching location will increase in two times. Their additional energy come in expense of electron energy from the second bunch, which are de-accelerated somewhat, while passing through the second bunch electrons. 

We can consider the injection of many identical electron bunches with the dwell time between them such that the previous bunches do not impact the dynamics of latter ones. One can easily find that the amount of such bunches is limited by $N_b\sim\ln(t/\tau_b)$. Therefore, maximum kinetic energy, acquired by the returned electrons of the very first bunch, after being accelerated
by electric field of all bunches can be estimated as $E_{k \max}\sim N_b\times E_{k\text{in}}$, which, nonetheless, can be
significantly larger than $E_{k\text{in}}$.

We can consider also continuous injection of electrons into a half-space taking the time-dependent distribution function of launching electrons $f(t,v)$. Considering the non-relativistic case we take,
\begin{equation}
f(t,v)=\frac{n_0\delta(v-v_0)}{1-\alpha t\omega_{pe}(n_0)},
\end{equation} 
where $\alpha\ll1$. This temporal evolution of electron launch, limited by $\alpha t\omega_{pe}(n_0)$, resembles the rate of bunch launches. The final energy by electric field of all bunches can be estimated as $N_b\times E_{k\text{in}}$, which, nonetheless, can be significantly larger than $E_{k\text{in}}$.

\begin{figure}
\includegraphics[width=8.5cm]{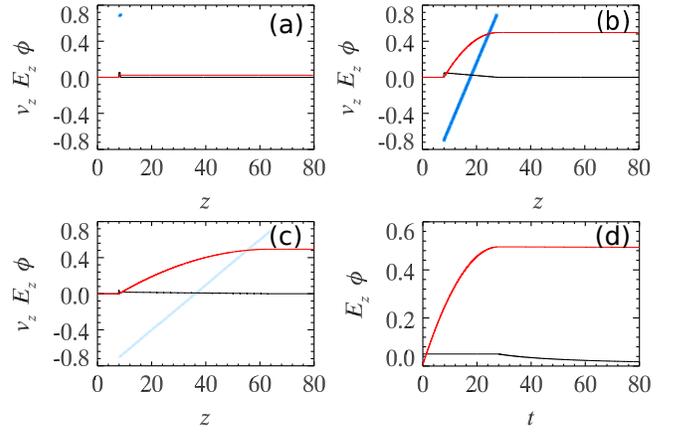}
\caption{\label{One_bunch_simulation} (color online) Simulation parameters, $t_b=L_0/v_0$, $t_p=1/\omega_{pe}$ and ${\tau}_b=2\tau_p^2/\tau_b$: $L_0=0.2$, $v_0=0.7$ and $\omega_{pe}=0.5$, corresponding to $t_b=0.28571$, $t_p=2.0$ and ${\tau}_b=28.0$. (a) (d) and (c) are the $z$-$v_z$ phase, electric field (black line) and potential (red line) profile at $t=0.5$, $t=28$ and $t=80$ respectively. (d) is the maximal electric field (black line) and potential (red line) evolution with time.}
\end{figure}

In order to confirm the above theoretical analysis, we also run a serious of 1-D electrostatic PIC simulations, which is solved by an energy conserving method (Appendix B). The electrostatic PIC simulations solve the following equations,
\begin{equation}
\frac{\partial f}{\partial t}+v\frac{\partial f}{\partial z}+\frac{eE}{m_e}\frac{\partial f}{\partial v}=0,
\end{equation}
\begin{equation}
\frac{\partial E}{\partial z}=4\pi e\int{f dv},
\end{equation}
\begin{equation}
f(t=0)=n_e\delta(v-v_0),
\end{equation}
with $\omega_{pe0}=4\pi n_0 e^2/m_e$, $v=\bar{v} [c]$, $t=\bar{t} [1/\omega_{pe0}]$, $z=\bar{z} [c/\omega_{pe0}]$, 
$E_{z}=\bar{E_{z}} [c\omega_{pe0}m_e/e]$, $\phi=\bar{\phi} [m_e c^2/e]$, $\omega_{pe}=\bar{\omega_{pe}} [\omega_{pe0}]$,
$n_e=\bar{\omega_{pe}}^2 [n_0]$ and $f=\bar{f} [n_0/c]$.
We define a reference density $n_0$, corresponding to an reference plasma frequency $\omega_{pe0}$. $1/\omega_{pe0}$ define the time scale in simulation, $c/\omega_{pe0}$ define the length scale and $c$ is speed of light. We can change the plasma density in simulation by adjusting $\bar{\omega_{pe}}$. If $\bar{\omega_{pe}}=1$, the plasma density used in simulation is exactly $n_0$, if $\bar{\omega_{pe}}=0.5$, the corresponding plasma density in simulation is $0.5\times0.5\times n_0$.

Fig.\ \ref{One_bunch_simulation} shows the simulation results, in which an electron bunch of velocity $v_0=0.7$, thickness $L_0=0.2$ and plasma frequency $\omega_{pe}=0.5$ is emitted from the surface $z=0$. Fig.\ \ref{One_bunch_simulation} (a), (b) and (c) show the time-snap of $z$-$v_z$ phase-space, electric field and potential profile at $t=0.5$, $t=28$ and $t=80$, which clearly demonstrates that at $t=28$, the electrons in the rear start returning to the emitting point at $z=0$, well consistent with the theoretical analysis, $\tau_d=(2/\omega^2_{pe})(v_0/L_0)=28$. In our simulation, we include a numerical friction mechanism to stopping electrons when re-entering into the emitting point. Fig. \ref{One_bunch_simulation} (d) shows the maximal electric field and potential evolution with time, and we find that the maximal potential almost keeps constant even when the back edge of the bunch returns to the emitting point, which is also consistent with theoretical prediction. As expected by theoretical analysis, the maximal electric field decrease with time as $\tau_b/t$ when $t>\tau_b$. 
The kinetic energy of the returned electron is exactly equal to the initial value, having $v=-v_0$, which is, nonetheless, consistent with the theoretical prediction.

\begin{figure}
\includegraphics[width=8.5cm]{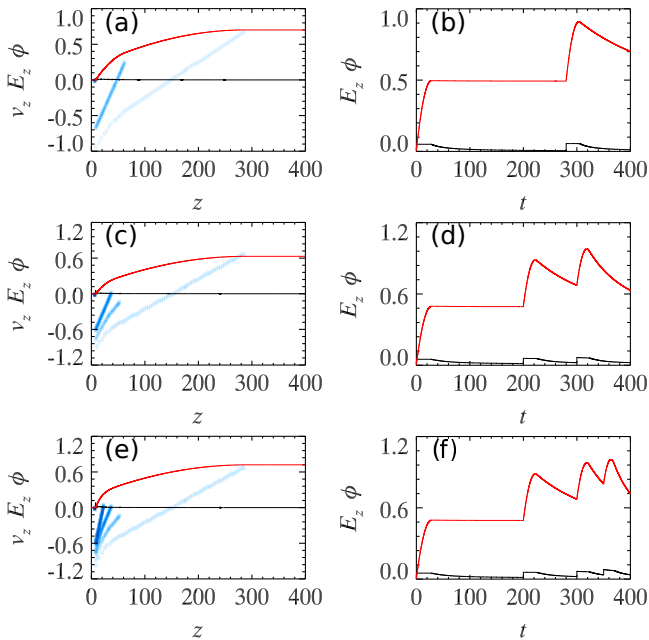}
\caption{\label{multi_bunch_simulation} (color online) Simulation parameters, $t_b=L_0/v_0$, $t_p=1/\omega_{pe}$ and ${\tau}_b=2\tau_p^2/\tau_b$: $L_0=0.2$, $v_0=0.7$ and $\omega_{pe}=0.5$, corresponding to $t_b=0.28571$, $t_p=2.0$ and ${\tau}_b=28.0$. (a)-(b) corresponds to two bunches cases, with the second bunch emitted at $t=200$. (c)-(d) corresponds to three bunches cases, with the third bunch emitted at $t=300$. (e)-(f) corresponds to four bunches cases, with the fourth bunch emitted at $t=350$. (a) (c) and (e) are the $z$-$v_z$ phase-space, electric field (black line) and potential (red line) profile at $t=400$ for two, three and four bunches cases respectively. (b) (d) and (f) are the corresponding maximal electric field (black line) and potential (red line) evolution with time.}
\end{figure}

Let us consider the situation of emitting multi bunches. Fig.\ \ref{multi_bunch_simulation} (a) and (b) show the two bunches cases with the dwell time $\tau_{\text{dw}}=280$ greatly larger than $\tau_{d}=28$. We noticed that the maximal potential energy can be further increased by the emission of the second bunch, finally reaching four times as large of original kinetic energy. 
The velocity of the returned electron can be as high as $v=-0.99$ compared with the initial value $v_0=0.7$, confirming the theoretical prediction that the kinetic energy of returned electron is increased by twice. 
Fig.\ \ref{multi_bunch_simulation} (c) (d) (e) and (f) are cases of three ($\tau_{\text{dw}1}=200$ and $\tau_{\text{dw}2}=100$) and four ($\tau_{\text{dw}1}=200$, $\tau_{\text{dw}2}=100$ and $\tau_{\text{dw}3}=50$) bunches, the maximal potential and the returned electron kinetic can be further increased as expected. Limited to the computational ability of our simulation, if the dwell time is long enough, the finial maximal potential energy will be close to the theoretically predicted value $E_{k \max}\sim\ln(t/\tau_b)E_{k\text{in}}$.

\begin{figure}
\includegraphics[width=8.5cm]{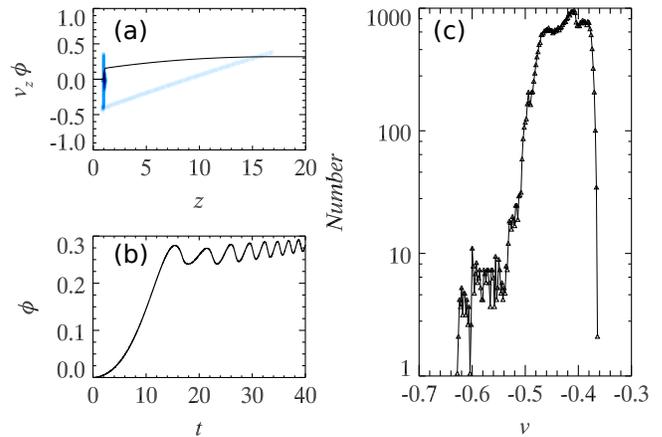}
\caption{\label{Continuous_bunch_simulation} (color online) Simulation parameters: electron beam with constant velocity $v_0=0.4$ and density profile $\omega_{pe}^2\exp{(t\omega_{pe})}$, where $\omega_{pe}=0.125$. (a) is the $z$-$v_z$ phase-space and potential profile plotted at $t=40$. (b) is the maximal potential evolution with time. (c) is the velocity spectra of the returned electrons collected at the emitting point.}
\end{figure}

As shown in Fig.\ \ref{fig3}, the emission of electrons is a continuous process. 
Here in Fig.\ \ref{Continuous_bunch_simulation}, we show the simulation results of continuous emission of electron beam with constant velocity $v_0=0.4$ and 
density profile $\omega_{pe}^2\exp{(t\omega_{pe})}$, where $\omega_{pe}=0.125$. The simulation results, as shown in Fig.\ \ref{Continuous_bunch_simulation} (b), indicate that the maximal potential energy is more than three times as large of the initial kinetic energy at $t=40$ and is still increasing gradually with time. Please note the oscillation of maximal potential energy, with its oscillation frequency increasing with time. These oscillations come from the plasma intrinsic oscillations, with its frequency determined by the density of the emitting electron beam. With the increase of density, the maximal potential energy and oscillation frequency are also increasing with time. Fig.\ \ref{Continuous_bunch_simulation} (c) records the velocity spectra of the returned electrons collected at the emission point, indicating that the returned electrons actually span a large velocity range, from $-0.3$ to $-0.7$. This spanning of the velocity range is also consistent with the theoretical prediction, with some of the electron having velocity higher than the initial value $0.4$, and some of the electron having velocity smaller than $0.4$. The cut-off kinetic energy of the returned electrons can be about three times as large of the initial value.

\section{Conclusions}
The generation of super-high energetic electrons influenced by pre-plasma in relativistic intensity laser matter interaction is studied in a one-dimensional slab approximation with particle-in-cell simulations. Different pre-plasma scale-lengths of $1\ \mu\text{m}$, $5\ \mu\text{m}$, $10\ \mu\text{m}$ and $15\ \mu\text{m}$ are considered, which shows an increase in both particle number and cut-off energy of energetic electrons with the increase of the pre-plasma scale-length, and the obtained cut-off energy of electrons greatly exceeding the corresponding laser ponderomotive energy. The two questions, i) ``why the generation efficiency of energetic electrons is increasing with the increase of pre-plasma scale-length'', and ii) ``what is underlying acceleration mechanism of super-high energetic electrons with kinetic energy greatly exceeding the ponderomotive energy'', are answered in this work.

A two-stage electron acceleration model is proposed to explain the underlying physics in detail. The first stage is attributed to the synergetic acceleration by the longitudinal charge separation electric field $E_z$ and the ponderomotive force of the reflected laser pulse. The efficiency of the first stage acceleration depends on the pre-plasma scale-length. 
The maximal possible energy gain during the first stage acceleration is analysed, and a scaling law is obtained by solving the relativistic electron motions in the presence of two counter-propagating plane laser waves and the external electric field due to the charge separation within limited space extension on the order of pre-plasma scale-length. 
The maximal-possible energy gain in the first stage is estimated to be $\varepsilon\ [\text{MeV}]=0.64\times \beta^{1/2} \times I^{1/2} \times L_p^{1/2}$, where $I$ is laser intensity normalized by $1.37\times10^{18}\ \text{W}/\text{cm}^2$ and $L_p$ is pre-plasma scale-length normalized by $\mu\text{m}$. The scaling law indicates that with the increase of pre-plasma scale-length and incident laser intensity, the maximal-possible electron energy is also increasing, which agrees well with the simulation results.

The energetic electrons pre-accelerated in the first stage could build up an intense electrostatic potential barrier with the potential energy several times as large of electron kinetic energy. Part of energetic electrons could be reflected by this potential, obtaining finial kinetic energies several times as large of the initial values.
The potential building and the accompanying electron kinetic enhancement process by this potential barrier are analysed theoretically and confirmed by electrostatic PIC simulations, where the theoretical prediction and electrostatic PIC simulations are in good agreement.

The multidimensional effects of laser propagation through under-dense plasmas are neglected in the present studies. We plan to address the multi-dimension effects in future studies.

\begin{acknowledgments}
This work was supported by the National Natural Science Foundation of China (11304331, 11174303, 61221064), the National Basic Research Program of China (2013CBA01504, 2011CB808104) and USDOE Grant DENA0001858 at UCSD.
\end{acknowledgments}

\begin{appendix}
\section{Confirmation of the reduced model}
\begin{figure}
\includegraphics[width=8.5cm]{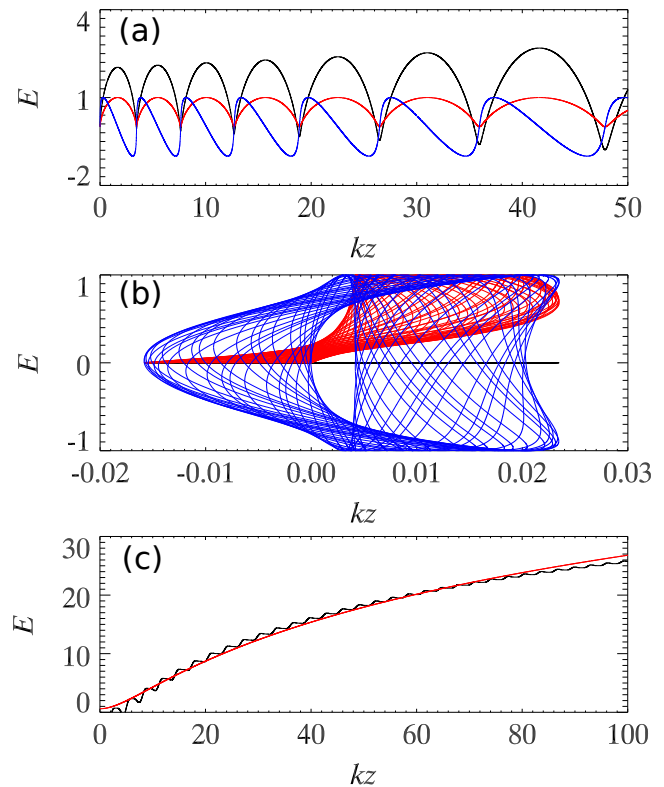}
\caption{\label{A1} (color online) Parameters: (a) $a_{+}(t-z)=2.0$, $a_{-}(t+z)=0.0$, $E_z=-0.02$, $p_z(t=0,z=0)=0.0$, 
(b) $a_{+}(t-z)=2.0$, $a_{-}(t+z)=2.0$, $E_z=-0.02$, $p_z(t=0,z=0)=0.0$ and (c) $a_{+}(t-z)=5.0$, $a_{-}(t+z)=5.0$, $E_z=-0.5$, $p_z(t=0,z=0)=10.0$. (a) and (b) black line represents the evolution of $\gamma_a\gamma_z-\gamma_a(t=0,z=0)\gamma_x(t=0,z=0)-E_z z$ vs $z$, red line represents $\sin(t-z)^2$ vs $z$ and blue line represents $\sin[2(t-z)]$ vs $z$. In (c) black line represents the evolution of $\gamma_a\gamma_z-\gamma_a(t=0,z=0)\gamma_x(t=0,z=0)-E_z z$ vs $z$ from full simulation, and red line represents the evolution of $\gamma_a\gamma_z-\gamma_a(t=0,z=0)\gamma_x(t=0,z=0)-E_z z$ vs $z$ from the reduced simulation.}
\end{figure}

We have studied the motion of a single electron in the field of $a_{+}$, $a_{-}$ and $E_z$ by numerically solving the $1$D-$3$V electron equation of motion with the standard Boris algorithm. Fig.\ \ref{A1} (a) shows the motion of a single electron in the fields of only $a_{+}$ and $E_z$. It indicates that when the Woodward-Lawson theorem is broken, electron will be continuously accelerated forward and the final kinetic energy is increasing with the increase of the acceleration length. Fig.\ \ref{A1} (b) shows when there exists two counter-propagating laser pulses, i.e. $a_{+}$ and $a_{-}$, the dynamics of the electron initially at rest is quite complicated, resulting in stochastic-like motions.  

However, when electron with a initial large momentum $p_z$ enters the fields of two counter-propagating laser waves and longitudinal electric field, the influence of the incident laser $a_{-}$ can be simplified. The only contribution of the incident laser wave $a_{-}$ is to increase the electron mass in an averaged way. 
In Fig.\ \ref{A1} (c), black line shows the full dynamics of the electron under $a_{+}$, $a_{-}$ and $E$, and the red line
shows the dynamics of electron only under $a_{+}$ and $E$ but replacing $\gamma_a=(1+a_{+}^2+a_{-}^2)^{1/2}$ to 
$\gamma_a=(1+a_{+}^2+a^2/2)^{1/2}$. The results of full dynamics and reduced model are well fitted, confirming our assumption.

\section{Simulation method of electrostatic PIC}

\end{appendix} 
A new PIC method, which conserves energy exactly, is used. The equations of motion of particles and the Maxwell's equations are
differenced implicitly in time by the mid-point rule and solved concurrently by a Jacobian-free Newton Krylov (JFNK) solver.
The particle average velocities and the electrostatic field are calculated self-consistently by the JFNK solver to preserve the total energy of the system.

{}


\begin{thebibliography}{99}

\bibitem{PhysRevLett.103.024801} M. Chen, A. Pukhov, T. P. Yu, and Z. M. Sheng, Phys. Rev. Lett. 103, 024801 (2009).

\bibitem{PhysRevLett.102.145002} B. Qiao, M. Zepf, M. Borghesi, and M. Geissler, Phys. Rev. Lett. 102, 145002 (2009).

\bibitem{PhysRevLett.103.135001} X. Q. Yan, H. C. Wu, Z. M. Sheng, J. E. Chen, and J. Meyer-ter Vehn, Phys. Rev. Lett. 103,
135001 (2009).

\bibitem{PhyPla.20.023012} D. Wu, C. Y. Zheng, C. T. Zhou, X. Q. Yan, M. Y. Yu, and X. T. He,
Phys. Plasmas 20, 023012 (2013).

\bibitem{PhyRevE.90.023101} D. Wu, C. Y. Zheng, B. Qiao, C. T. Zhou, X. Q. Yan, M. Y. Yu, and X. T. He,
Phys. Rev. E 90, 023101 (2014).

\bibitem{PhysRevLett.115.064801} J. H. Bin, W. J. Ma, H. Y. Wang, M. J. V. Streeter, C. Kreuzer, D. Kiefer, M. Yeung, S. Cousens, P. S. Foster, B. Dromey, X. Q. Yan, R. Ramis, J. Meyer-ter-Vehn, M. Zepf, and J. Schreiber,
Phys. Rev. Lett. 115, 064801 (2015).

\bibitem{PhyPla.20.013101} H. Y. Wang, X. Q. Yan, J. E. Chen, X. T. He, W. J. Ma, J. H. Bin, J. Schreiber, T. Tajima, 
and D. Habs, Phys. Plasmas 20 013101 (2013).
\bibitem{PhyPla.1.1626} M. Tabak, J. Hammer, M. E. Glinsky, W. L. Kruer, S. C. Wilks, J. Woodworth, E. M. Campbell, 
M. D. Perry, and R. J. Mason, Phys. Plasmas 1, 1626 (1994).

\bibitem{PhyPla.15.056304} L. Van Woerkom, K. U. Akli, T. Bartal, F. N. Beg, S. Chawla, C. D. Chen, E. Chowdhury, R. R. Freeman, D. Hey, M. H. Key, J. A. King, A. Link, T. Ma, A. J. MacKinnon, A. G. MacPhee, D. Offermann, V. Ovchinnikov, P. K. Patel, 
D. W. Schumacher, R. B. Stephens, and Y. Y. Tsui, Phys. Plasmas 15, 056304 (2008).

\bibitem{PhysRevLett.104.055002} A. G. MacPhee, L. Divol, A. J. Kemp, K. U. Akli, F. N. Beg, C. D. Chen,
H. Chen, D. S. Hey, R. J. Fedosejevs, R. R. Freeman, M. Henesian, M. H. Key, S. Le Pape, A. Link, T. Ma, 
A. J. Mackinnon, V. M. Ovchinnikov, P. K. Patel, T. W. Phillips, R. B. Stephens, M. Tabak, R. Town, Y. Y. Tsui,
L. D. Van Woerkom, M. S. Wei, and S. C. Wilks, Phys. Rev. Lett. 104, 055002 (2010).

\bibitem{PhysRevLett.108.115004} T. Ma, H. Sawada, P. K. Patel, C. D. Chen, L. Divol, D. P. Higginson, A. J. Kemp, M. H. Key, D. J. Larson, S. Le Pape, A. Link, A. G. MacPhee, H. S. McLean, Y. Ping, R. B. Stephens, S. C. Wilks, and F. N. Beg, 
Phys. Rev. Lett, 108, 115004 (2012).
\bibitem{Nat.Phys.7.867} S. Cipiccia, M. R. Islam, B. Ersfeld, R. P. Shanks, E. Brunetti, G. Vieux,
X. Yang, R. C. Issac, S. M. Wiggins, G. H. Welsh, M.-P. Anania, D. Maneuski, R. Montgomery, G. Smith, M. Hoek, D. J. Hamilton, N. R. C. Lemos, D. Symes, P. P. Rajeev, V. O. Shea, J. M. Dias, and D. A. Jaroszynski, Nat. Phys. 7, 867–871 (2011).

\bibitem{PhyPla.22.080704} B. Liu, R. H. Hu, H. Y. Wang, D. Wu, J. Liu, C. E. Chen, J. Meyer-ter-Vehn, X. Q. Yan, and X. T. He,
Phys. Plasmas 22, 080704 (2015).
\bibitem{PhyPla17.060704} T. Yabuuchi, B. S. Paradkar, M. S. Wei, J. A. King, F. N. Beg, R. B. Stephens, N. Nakanni, 
H. Hatakeyama, H. Habara, K. Mima, K. A. Tanaka, and J. T. Larsen, Phys. Plasmas 17, 060704 (2010).

\bibitem{Phys.Rev.E.79.066406} A. J. Kemp, Y. Sentoku, and M. Tabak, Phys. Rev. E 79, 066406 (2009).

\bibitem{PhyPlas.17.023106} H. Cai, K. Mima, A. Sunahara, T. Johzaki, H. Nagatomo, S. Zhu, and X. T. He, 
Phys. Plasmas 17, 023106 (2010).

\bibitem{PhyPlas.16.103101} M. Sherlock, Phys. Plasmas 16, 103101 (2009).

\bibitem{Phys.Rev.E.83.046401} B. S. Paradkar, M. S. Wei, T. Yabuuchi, R. B. Stephens, M. G. Haines, S. I. Krasheninnikov, and F. N. Beg, Phys. Rev. E, 83, 046401 (2011).

\bibitem{PhyPla.20.012703} Wei-wu Wang, Hong-bo Cai, Qing Jia, and Shao-ping Zhu, Phys. Plasmas 20, 012703 (2013).

\bibitem{App.Phys.Lett.102.224101} W. P. Wang, B. F. Shen, H. Zhang, Y. Xu, Y. Y. Li, X. M. Lu, C. Wang, Y. Q. Liu, J. X. Lu, Y. Shi, Y. X. Leng, X. Y. Liang, R. X. Li, N. Y. Wang, and Z. Z. Xu, App. Phys. Lett. 102, 224101 (2013).

\bibitem{Phys.Fluids.30.526} G. Sun, Ott E, Y. Lee, and P. Guzdar, Phys. Fluids 30, 526 (1987).

\bibitem{Phys.Rev.Lett.33.209} C. E. Max, J. Arons, and A. B. Langdon, Phys. Rev. Lett. 33, 209 (1974).

\bibitem{J.Comp.Phys.227.6846} Y. Sentoku and A. J. Kemp, Journal of Computational Physics 227, 6846 (2008).

\bibitem{PhyPlas.22.093108} D. Wu, B. Qiao, and X. T. He, Phys. Plasmas 22, 093108 (2015).

\bibitem{IEE.T.N.S.26.4217} J. D. Lawson, IEE Trans. Nucl. Sci. 26, 4217 (1979).

\bibitem{PhyPla.19.060703} B. S. Paradkar, S. I. Krasheninnikov, and F. N. Beg, Phys. Plasmas 19, 060703 (2012).

\bibitem{Phy.Rev.Lett.111.065002} A. P. L. Robinson, A. V. Arefiev, and D. Neely, Phys. Rev. Lett. 111, 065002 (2013).

\bibitem{PhyPla.21.104510} S. I. Krasheninnikov, Phys. Plasmas 21, 104510 (2014). 

\end{thebibliography}
\end{document}